\documentclass[aps,prl,twocolumn,showpacs,superscriptaddress,groupedaddress]{revtex4-1}

\usepackage{bm}
\usepackage[usenames,dvipsnames]{color}
\usepackage{dcolumn}
\usepackage[usenames,dvipsnames,svgnames,table]{xcolor}
\usepackage{amssymb}
\usepackage{amsmath}
\usepackage{bbold}
\usepackage{amsfonts}
\usepackage{graphicx}
\usepackage{xstring}
\usepackage{units}

\newcommand{\tauon}{\tau_{\mathrm{on}}}
\newcommand{\D}{\mathit{\Delta}}
\renewcommand{\v}[1]{\boldsymbol{#1}}
\newcommand{\wo}[1]{_{\backslash #1}}
\newcommand{\T}{^\mathsf{T}}
\newcommand{\cp}[3][]{p\IfStrEqCase{#1}{{}{}}[#1](#2\,|\,#3)}
\newcommand{\normal}[2][]{\mathcal N(#2 \IfStrEqCase{#1}{{}{}}[;\,#1])}
\newcommand{\expect}[2][]{\left\langle \, #2\, \right\rangle_{#1} }

\newcommand{\tauref}{\tau_\mathrm{ref}}
\newcommand{\uthr}{\vartheta}  

\newcommand{\ueff}{u_\mathrm{eff}}
\newcommand{\taueff}{\tau_\mathrm{eff}}

\newcommand{\Erev}{E^\mathrm{rev}}

\newcommand{\El}{E_\mathrm{l}}
\newcommand{\gl}{g_\mathrm{l}}
\newcommand{\Cm}{C_\mathrm{m}}
\newcommand{\taum}{\tau_\mathrm{m}}

\newcommand{\tausyn}{\tau_\mathrm{syn}}

\newcommand{\gtot}{g_\mathrm{tot}}

\newcommand{\Isyn}{I^\mathrm{syn}}

\newcommand{\Irec}{I^\mathrm{rec}}
\newcommand{\Iext}{I^\mathrm{ext}}
\newcommand{\Inoise}{I^\mathrm{noise}}
\newcommand{\wnoise}{w^\mathrm{noise}}
\newcommand{\gnoise}{g^\mathrm{noise}}

\newcommand{\erf}{\mathrm{erf}}
\newcommand{\thetaeff}{\vartheta_\mathrm{eff}}

\newcommand{\Dkl}[2]{\textnormal{D}_\textnormal{KL}\left( #1 \,||\, #2 \right)}

\begin{document}

\preprint{APS/123-QED}

\title{Stochastic inference with deterministic spiking neurons}
\thanks{
The first two authors contributed equally to this work.
We thank W.~Maass for his essential support and Stefan Habenschuss for helpful comments.
This research was supported by EU grants \#269921 (BrainScaleS), \#237955 (FACETS-ITN), the Austrian Science Fund FWF \#I753-N23 (PNEUMA) and the Manfred St\"ark Foundation.
}

\author{Mihai A. Petrovici$^1$*, Johannes Bill$^2$*, Ilja Bytschok$^1$, Johannes Schemmel$^1$, Karlheinz Meier$^1$}
\affiliation{$^1$Kirchhoff Institute for Physics, University of Heidelberg \\ $^2$Institute for Theoretical Computer Science, Graz University of Technology}

\date{\today}

\begin{abstract}

    The seemingly stochastic transient dynamics of neocortical circuits observed  \textit{in vivo} have been hypothesized to represent a signature of ongoing stochastic inference.
    \textit{In vitro} neurons, on the other hand, exhibit a highly deterministic response to various types of stimulation.
    We show that an ensemble of deterministic leaky integrate-and-fire neurons embedded in a spiking noisy environment can attain the correct firing statistics in order to sample from a well-defined target distribution.
    We provide an analytical derivation of the activation function on the single cell level; for recurrent networks, we examine convergence towards stationarity in computer simulations and demonstrate sample-based Bayesian inference in a mixed graphical model.
    This establishes a rigorous link between deterministic neuron models and functional stochastic dynamics on the network level.

\end{abstract}

\pacs{xxx-xxx}
\maketitle

\subsection{Introduction}

In responding to environmental sensory stimuli, brains have to deal with what is typically limited, noisy and ambiguous data.
Based on such imperfect information, animals need to predict and react to changes in their environment.
The recent hypothesis that the brain handles these challenges by performing Bayesian, rather than logical inference \cite{kording2004bayesian, fiser2010statistically, friston2011action}, has been strengthened by electrophysiological data which identified neural correlates of the involved computations \cite{yang2007probabilistic,berkes2011spontaneous} and theoretical work on spiking network implementations \cite{deneve2008bayesian, buesing2011neural, rao2004hierarchical}.

In probabilistic inference, the potential values of a quantity are described by a random variable (RV) $z_k$
and all knowledge about dependencies between random variables is stored in a joint probability distribution
$p(z_1,\dots,z_K)$.
The Bayesian belief about a set of unobserved RVs $\{z_1, \dots, z_M\}$ given an observed set of RVs is represented by the posterior distribution
$\cp{z_1,\dots,z_M}{z_{M+1},\dots,z_K}$.
In particular, the posterior contains information not only on the most likely conclusion and potential alternatives, but also on the level of uncertainty of the outcome.

Theoretical work \cite{fiser2010statistically} has argued in favor of sample-based representations of probability distributions in the brain.
In this representation, instead of providing the entire distribution at any point in time, only samples \mbox{$\v z^{(t)} \sim p(z_1,\dots,z_K)$} are used as a proxy.
When modeling large systems, this offers three important advantages.
First, approximate solutions can be provided at any time, with increasingly reliable results as the calculation progresses
(``anytime computing'').
Secondly, in a sample-based representation marginalization comes at no cost, as $p(z_k)$ can be determined by simply neglecting the values of all other RVs.
Thirdly, some sampling algorithms such as Gibbs sampling support a high degree of parallelization.
In particular, the resulting algorithmic structure is reminiscent of neural networks \cite{hoyer2003interpreting}.

\begin{figure}[tbp]
    \centering
    \includegraphics[]{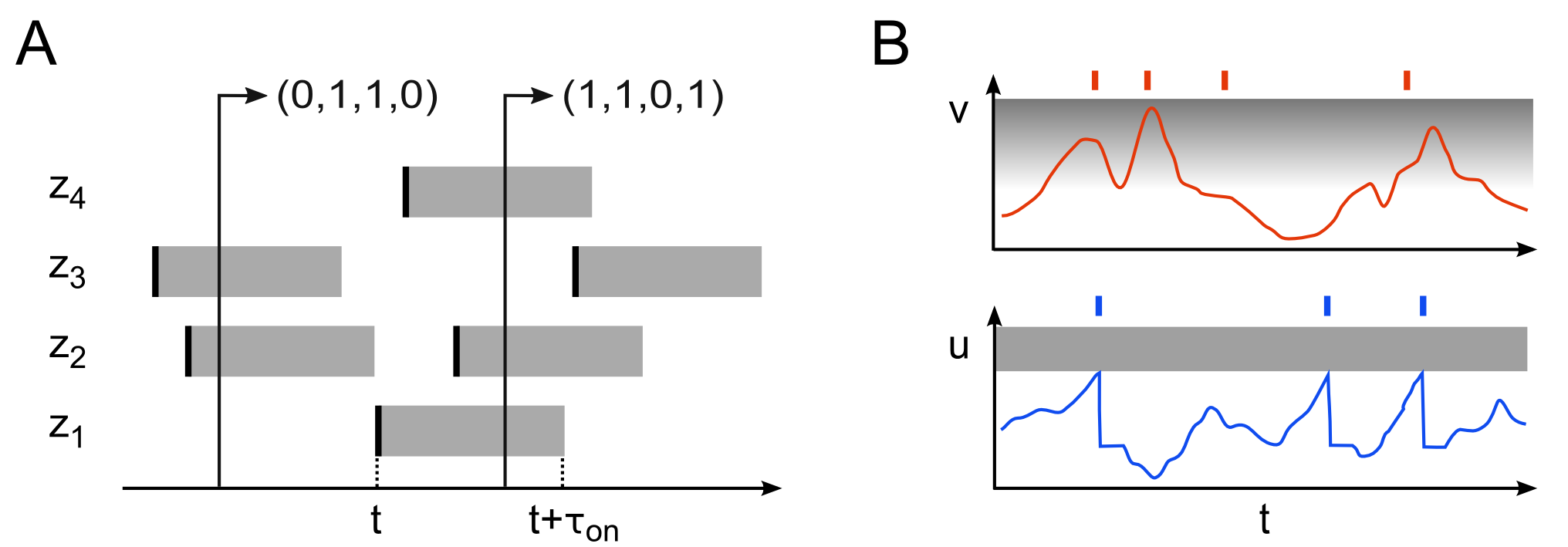}
    \caption{
        (A) Interpretation of spike patterns as samples of a binary random vector $\v z$.
        The variable $z_k$ is active for duration $\tauon$ (grey bar) after a spike of neuron $k$.
        (B) In stochastic neuron models, internal state variables modulate the instantaneous firing probability (red).
        In contrast, deterministic integrate-and-fire neurons elicit a spike when the membrane potential crosses a threshold voltage (blue).
        The probability of firing as a function of the respective internal variable is represented by the grayscale in the background.
        \label{fig:1}
    }
\end{figure}

Recently, a theory has been suggested which combines these advantages by implementing Markov chain Monte Carlo sampling in networks of abstract model neurons \cite{buesing2011neural}.
In this framework, spike patterns are interpreted as samples of binary RVs as follows (see Fig.~\ref{fig:1}A):
\begin{align}
  z_k^{(t)} = 1 & \Leftrightarrow \mathrm{Neuron} \; k \; \mathrm{fired \; in} \; (t-\tauon, t] \quad .
\end{align}
The duration $\tauon$ of the active state following a spike is a free parameter; in cortex, $\tauon \approx \unit[10]{ms}$ is a good estimate for the timescale on which a spiking neuron affects the membrane potential of downstream cells.
The neuron model underlying \cite{buesing2011neural} is inherently stochastic (Fig.~\ref{fig:1}B top), with an instantaneous firing rate defined by
\begin{align}
    r_k(t) = \lim_{\D t \to 0} \frac{p(\mathrm{spike \; in \;} [t,\,t+\D t))}{\D t} \nonumber \\
    = \left\{
    \begin{array}{ll} 
        \frac{1}{\tau} \exp(v_k) & \; \mathrm{if} \; z_k = 0 \\
        0 \quad\;\; & \; \mathrm{if} \; z_k=1 \quad ,
    \end{array} 
    \right.
  \label{eqn:stochastic_neurons}
\end{align}
where $v_k$ represents an abstract membrane potential.

In contrast to this approach, \textit{in vitro} experiments have demonstrated the largely deterministic nature of single neurons \cite{mainen1995reliability}.
Similarly, microscopic models of neural circuits typically rely on deterministic dynamics of their constituents.
An often-used mechanistic model is the leaky integrate-and-fire (LIF) neuron
\begin{equation}
    \Cm\, \frac{du_k}{dt} = \gl (\El - u_k) + I \quad ,
    \label{eqn:LIF_dynamics}
\end{equation}
with capacitance $\Cm$, membrane potential $u_k$, leak potential $\El$, leak conductance $\gl$ and input current $I$.
The spiking condition is deterministic as well: when $u_k$ crosses a threshold $\uthr$ from below, a spike is emitted and $u_k$ is  reset to $\varrho$ for a refractory period $\tauref$ (see Fig.~\ref{fig:1}B bottom).
For conductance-based synapses, the synaptic input current to a neuron is typically modelled as
\begin{equation}
    \frac{d\Isyn_k}{dt} = -\frac{\Isyn_k}{\tausyn} + \sum_{\mathrm{syn} \; i} \sum_{\mathrm{spk} \; s} w_{ki} \left( \Erev_i - u_k \right) \delta(t-t_s) \; ,
    \label{eqn:IsynODE_maintxt}
\end{equation}
with the synaptic time constant  $\tausyn$, the synaptic weight $w_{ki}$ and the reversal potential of the $i$th synapse $\Erev_i$.

The aim of this letter is to demonstrate how a network of deterministic neurons in a biologically plausible spiking noisy environment can quantitatively reproduce the stochastic dynamics required for sampling from a well-defined distribution $p(z_1,\dots,z_K)$ and perform inference given observations.
We start by calculating the dynamics of a single LIF neuron in a spiking noisy environment and derive its activation function by describing the spike response as a first passage time (FPT) problem.
This establishes an equivalence to the abstract, inherently stochastic units (\ref{eqn:stochastic_neurons}).
On the network level, we show how biologically realistic conductance-based synapses (\ref{eqn:IsynODE_maintxt}) approximate the interaction for sampling from a well-defined target distribution.
We complement our study with a demonstration of probabilistic inference by implementing the posterior of a small graphical model for handwritten digit recognition in a recurrent network of LIF neurons.

\subsection{Deterministic neurons in a noisy environment}

The total input current $I_k$ to a neuron can be formally partitioned into recurrent synaptic input, diffuse synaptic noise and additional external currents:
$I_k = \Irec_k + \Inoise_k + \Iext_k$.
While the synaptic currents $\Irec_k$ and $\Inoise_k$ obey eqn.~(\ref{eqn:IsynODE_maintxt}),
the current $\Iext_k$ captures additional current stimuli.
We start by considering a single neuron that receives diffuse synaptic noise $\Inoise_k$ in the form of random spikes from its surrounding.
The capacity of recurrent networks to produce such noise has been shown in \cite{brunel2000dynamics}.
Throughout the following analysis of individual neurons we omit the index $k$ and set $\Irec = 0$.

When a conductance-based LIF neuron receives strong synaptic stimulation, it enters a so-called high-conductance state (HCS, \cite{destexhe03hcs}), characterized by accelerated membrane dynamics.
It is therefore convenient to rewrite (\ref{eqn:LIF_dynamics}) as
\begin{equation}
    \taueff \frac{du}{dt} = \ueff - u \quad ,
    \label{eqn:reduceddudt_maintext}
\end{equation}
where the membrane time constant $\taum = \Cm / \gl$ is replaced by a smaller effective time constant \mbox{$\taueff = \Cm/\gtot$}, with the total conductance $\gtot$ subsuming both leakage and synaptic conductances.
In a HCS, $\taueff$ governs the decay towards an effective leak potential $\ueff = (\gl\El + \sum_i \gnoise_i \Erev_i + \Iext)/\gtot$, where $\gnoise_i$ represents the total conductance at the $i$th synapse.
In a high input rate regime, $\sqrt{\mathrm{Var}(\gtot)}/\expect{\gtot} \rightarrow 0$ and the equation governing the membrane potential can be written as
\begin{equation}
    \taueff \frac{du}{dt} = \frac{\Iext + \gl\El}{\expect{\gtot}} + \frac{\sum_i \gnoise_i \Erev_i}{\expect{\gtot}} - u \quad ,
\end{equation}
with $\expect{\cdot}$ denoting the mean.
In a first approximation, $\taueff$ can be considered very small in the HCS, resulting in $u \approx \ueff$, with the effective potential $\ueff$ simply being a linear transformation of the synaptic noise input.

Using methods similar to \cite{ricciardi1979ouprocess}, it can be shown that, if stimulated by a large number of uncorrelated spike sources, the synaptic current $\Inoise$ -- and therefore, also $\ueff$ -- can be described as an Ornstein-Uhlenbeck (OU) process:
\begin{equation}
    du(t) = \theta \cdot (\mu - u(t)) + \mathit{\Sigma} \cdot dW(t) \; ,
\end{equation}
with parameters
\begin{align}
    & \theta = \tausyn \\
    & \mu = \frac{\Iext + \gl\El + \sum_i \nu_i \wnoise_i \Erev_i \tausyn}{\expect{\gtot}} \\
    & \mathit{\Sigma}^2 = \sum_i \nu_i \left[ \wnoise_i \left( \Erev_i - \mu \right) \right]^2 \tausyn \; / \expect{\gtot} \quad .
\end{align}
where $\nu_i$ represents the input rate at the $i$th noise synapse and $w_i$ its  weight.

\subsection{The activation function as an FPT problem}

The inherently stochastic neuron model (\ref{eqn:stochastic_neurons}) leads to a logistic activation function for constant potential $v$:
\begin{align}
    p(z = 1) &= \sigma(v) :=\left[ 1+\exp(-v) \right]^{-1} \quad .
    \label{eqn:activesigma}
\end{align}
In the following, we derive the activation function of the deterministic LIF neuron in a spiking noisy environment.
Similarly to the abstract model \cite{buesing2011neural}, we define the refractory state of a neuron as $z = 1$.

\begin{figure}[!ht]
    \centering
    \includegraphics[]{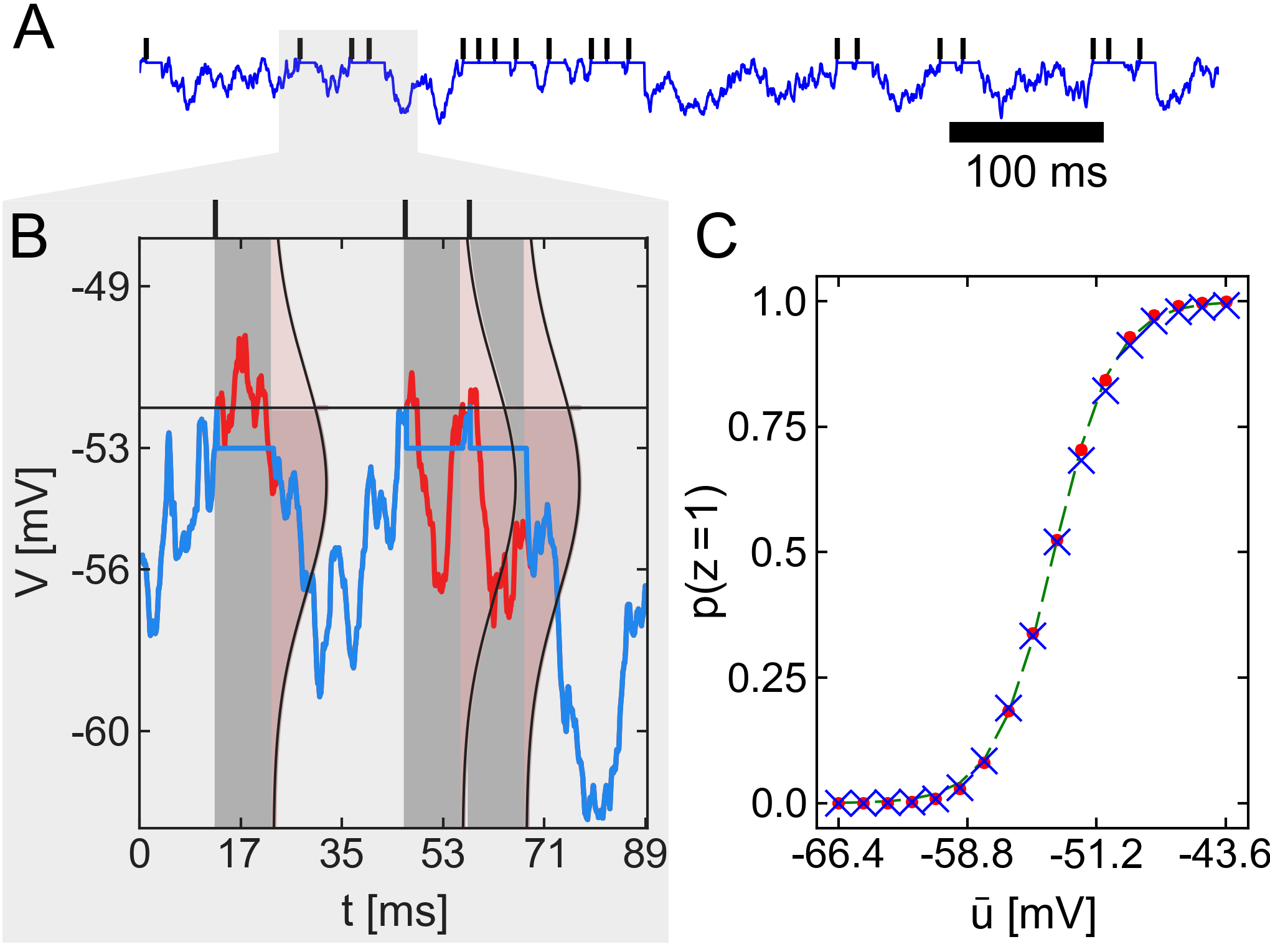}
    \caption{
        (A) Membrane potential $u(t)$ (blue) and resulting spike activity (black) of a LIF neuron in a spiking noisy environment.
        (B) Overlay of $u$ (blue) and $\ueff$ (red).
        Due to the small $\taueff$ in the HCS, the two curves are nearly identical when the neuron is not refractory.
        At the end of each refractory state (corresponding to $z=1$, grey), the predicted probability distribution for $\ueff$ is plotted in pink.
        The normalized subthreshold area (dark pink) is used for the propagation in (\ref{eqn:fullrecursion1_maintext}).
        (C) Theoretical prediction (red) vs. simulation results (blue); errors are smaller than the symbol size.
        A logistic function $\sigma(\bar u)$ (green) has been fitted to the prediction. 
        \label{fig:2}
        }
\end{figure}

An example of membrane potential dynamics with reset is shown in Fig.~\ref{fig:2}A.
Two modes of firing can be observed: the ``bursting'' mode, where the effective membrane potential after the refractory period is still above threshold, and the freely evolving mode, where the neuron does not spike again immediately after the refractory period.
Denoting the relative occurrence of burst lengths $n$ by $P_n$ and the average duration of the freely evolving mode that follows an $n$-spike-burst by $T_n$, we can identify the following relation:
\begin{equation}
    p(z = 1) = \frac{\sum_n P_n \cdot n \cdot \tauon}{\sum_n P_n \cdot (n \cdot \tauon + T_n)} \quad .
    \label{eqn:activationburstsum_maintext}
\end{equation}
Given the parameters of the associated OU process, we can derive a recursive expression for $P_n$ and $T_n$, thereby ultimately allowing the calculation of $p(z = 1)$:
\begin{align}
    P_n & = p(u_n < \uthr , u_{n-1} \geq \uthr, \dots, u_1 \geq \uthr | u_0 = \uthr) \label{eqn:fullrecursion1_maintext} \\
        & = \left( 1 - \sum_{i=1}^{n-1} P_i \right) \int_{\uthr}^\infty du_{n-1} p(u_{n-1} | u_{n-1} \geq \uthr) \nonumber \\
		& \quad \left[ \int_{-\infty}^{\uthr} du_n p(u_n | u_{n-1}) \right] \quad , \nonumber \\
    T_n & = \int_{\uthr}^\infty du_{n-1} p(u_{n-1} | u_{n-1} \geq \uthr) \label{eqn:fullrecursion2_maintext}\\
		& \quad \left[ \int_{-\infty}^{\uthr} du_n p(u_n | u_n < \uthr, u_{n-1}) \expect{T(\uthr, u_n)} \right] \quad . \nonumber
\end{align}
Fig.~\ref{fig:2}B displays an intuitive picture of the integrals in (\ref{eqn:fullrecursion1_maintext}) and (\ref{eqn:fullrecursion2_maintext}).
The transfer function $p(u_n | u_{n-1})$ is the Green's function of the OU process for $t=\tauref$:
\begin{equation}
    p(u_n|u_{n-1}) = C \, e^{-\frac{\theta}{\mathit{\Sigma}^2}\left[\frac{(u_n - (u_{n-1} - \mu) \exp(-\theta \tauref) - \mu)^2 }{1-\exp(-2\theta \tauref)}\right]} \; ,
\end{equation}
with the normalization $C=\sqrt{\theta / \pi \mathit{\Sigma}^2 (1-e^{-2\theta \tauref})}$.
$T(u_b, u_a)$ denotes the time the membrane needs to reach $u_b$ starting from $u_a$. 
This FPT problem has been extensively discussed in literature and its moments can be given in closed form \cite{ricciardi1988fptdensity}.

To improve the prediction of the activation function, we further take into account small, but finite $\taueff$, in which case the membrane potential no longer directly follows the input current, but is a low-pass-filtered version thereof. By using an expansion in $\sqrt{\taueff/\tausyn}$, a first-order correction to the FPT can be calculated \cite{brunel1998firingfrequency}:
\begin{equation}
  \expect{T(\vartheta, u)} = \tausyn \sqrt{\pi} \int\limits_{\frac{u - \mu}{\sigma}}^{\frac{\thetaeff-\mu}{\sigma}} dx \exp(x^2)[\erf(x) + 1] \;\; ,
  \label{eqn:newpassagetime_maintext}
\end{equation}
with the effective threshold $\thetaeff \approx \uthr - \zeta\left(\frac{1}{2}\right) \sqrt{\frac{\taueff}{2\tausyn}}$, where $\zeta$ denotes the Riemann Zeta function.
A comparison of the predicted $p(z=1)$ with results from a numerical simulation is shown in Fig.~\ref{fig:2}C.
Here, the average effective potential $\bar u = \expect{\ueff}$ was established through an external current $\Iext$.
For the translation from the LIF domain to the abstract stochastic model (\ref{eqn:stochastic_neurons}) we identify
\begin{align}
    v &= \frac{\bar u - \bar u^0}{\alpha} \quad ,
    \label{eq:linear_trans_membrane}
\end{align}
where $\bar u^0$ denotes the value of $\bar u$ for which \mbox{$p(z=1) = \sigma(0) = \frac{1}{2}$ and $\alpha$} represents a scaling factor between the two domains.
We conclude that a single LIF neuron in a spiking noisy environment can closely reproduce the activation function (\ref{eqn:activesigma}).

\subsection{Sampling via recurrent networks of LIF neurons with conductance-based synapses}

We next connect the neurons to a recurrent network.
In addition to noise stimuli, a LIF neuron receives synaptic currents $\Irec_k$ from other neurons in the ensemble.
Synaptic interaction introduces correlations among the spike response of different neurons, i.e.\ the random variables $z_1,\dots,z_K$ are not independent.
For certain connectivity structures, it is possible to specify a target distribution \cite{buesing2011neural,pecevski2011probabilistic} for the network states $\v z^{(t)}$ that occur under the network dynamics .

In the following, we use the emulation of Boltzmann machines as an example case.
The joint distribution reads:
\begin{align}
  \label{eq:BM_joint}
  p_\mathrm{B}(\v z) &= \frac{1}{Z}\,\exp\left(\frac12 \v z\T \v W \v z + \v z\T \v b \right) \quad ,
\end{align}
where $\v W$ is a symmetric zero-diagonal weight matrix, $\v b$ is a bias vector, $\v z\T \v W \v z$ and $\v z\T \v b$ denote bilinear forms over $\v W$ and $\mathbb{1}$ respectively, and $Z = \sum_{\v z'} \exp\left(\frac12 \v {z'}\T \v W \v z' + \v {z'}\T \v b \right)$ is the partition function that ensures correct normalization.
This probabilistic model underlies state-of-the-art machine learning algorithms for image \cite{salakhutdinov2009deep} and speech recognition \cite{mohamed2012acoustic}.
For the abstract neuron model (\ref{eqn:stochastic_neurons}), it had been proven \cite{buesing2011neural} that a membrane potential of the form
\begin{align}
  \label{eq:BM_vmem}
  v_k &= b_k + \sum_{j=1}^K W_{kj}\,z_j
\end{align}
leads to the desired target distribution (\ref{eq:BM_joint}) of network states $\v z^{(t)}$ for $t \rightarrow \infty$.
This finding uses the fact that individual neurons can sample from the conditionals \mbox{$\cp{z_k = 1}{\v z \wo k} = \sigma(v_k)$}, with $\v z \wo k = \{ z_j \, | \, j \neq k \}$, in a Gibbs sampling inspired updating scheme.

As shown above, LIF neurons in a spiking noisy environment closely approximate this logistic activation function if the synaptic currents $\Irec$ shift the mean membrane potential $\bar u_k$ according to the linear interaction (\ref{eq:BM_vmem}).
Using the linear transformation (\ref{eq:linear_trans_membrane}) between $v_k$ and $\bar u_k$, and estimating the impact of a pre-synaptic spike on the post-synaptic neuron through conductance-based synapses of weight $w_{kj}$, we arrive at the following parameter translation between the abstract and the LIF domain:
\begin{align}
    b_k = & (\bar u^b_k - \bar u^0_k) / \alpha \label{eqn:biastrans}\\
    W_{kj} = & \frac{1}{\alpha \Cm} \frac{w_{kj} \left(\Erev_{kj} - \mu\right)}{1 - \frac{\tausyn}{\taueff}} \nonumber \\
            & \left[\tausyn \left( e^{-1} - 1 \right) - \taueff \left( e^{- \frac{\tausyn}{\taueff}} - 1 \right) \right] \quad,
    \label{eq:weight_trans}
\end{align}
where $\bar u^b_k$ is the mean free potential $\bar u_k$ in Fig.~\ref{fig:2}C that establishes $\cp{z_k = 1}{\v z \wo k = \v 0} = \sigma(b_k)$, and $\Erev_{kj}$ denotes the reversal potential for synapse $w_{kj}$.
The idea behind (\ref{eq:weight_trans}) is to match the integrals of individual postsynaptic potentials (PSPs) on $v_k$ and $\bar u_k$.
We furthermore employ short-term synaptic depression to approximate the theoretically optimal rectangular PSP shape also in case of consecutive spikes (bursts).

\begin{figure}[!t]
    \centering
    \includegraphics[width=8.6cm]{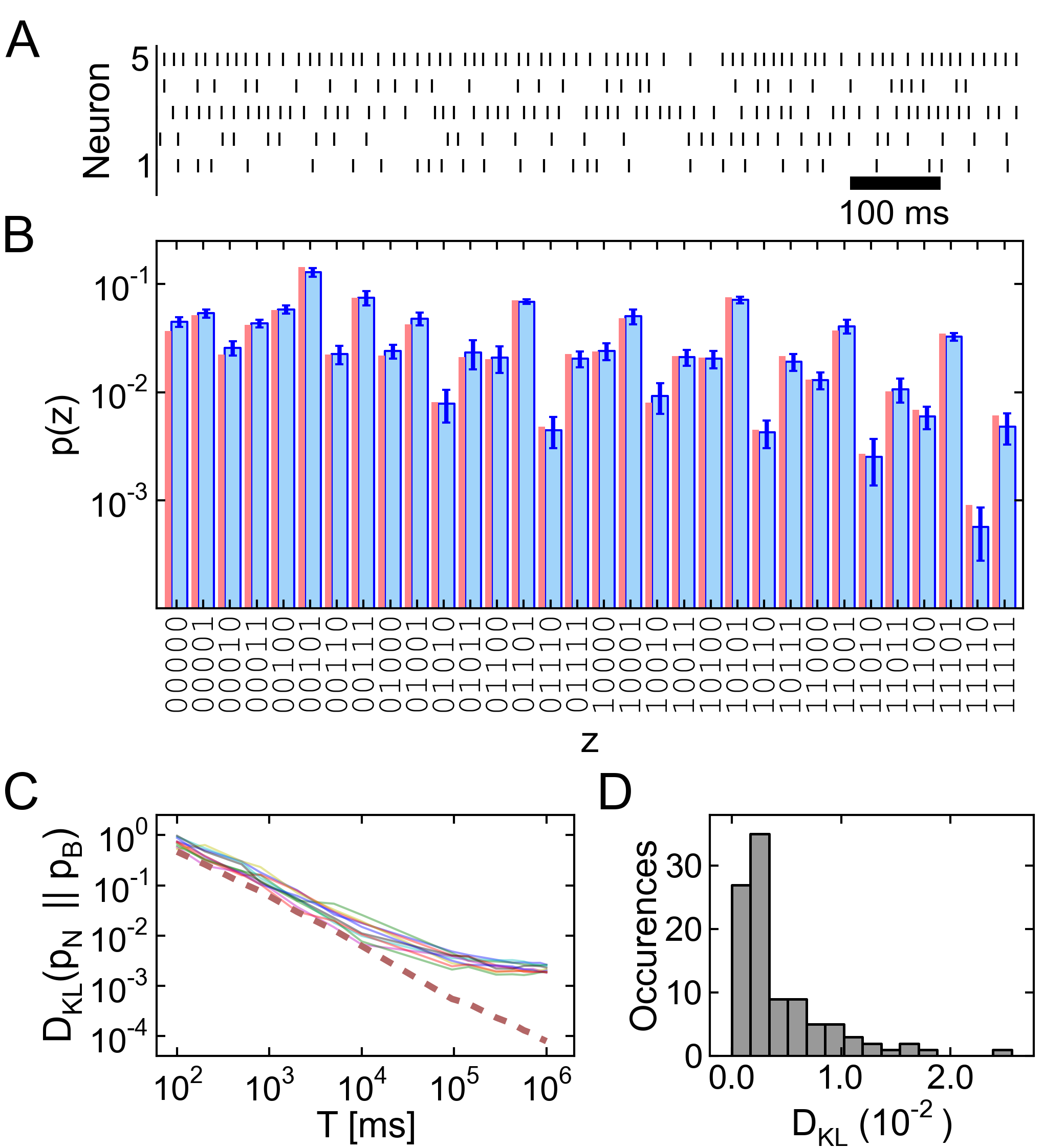}
    \caption{\label{fig:3}
        (A) Spike pattern of a recurrent network of LIF neurons during sampling from a randomly generated Boltzmann machine.
        (B) Sampled distribution $p_\mathrm{N}(\v z)$ of network states (blue bars) and analytically calculated target distribution $p_\mathrm{B}(\v z)$ (in red).
        (C) Kullback-Leibler divergence between the sampled distribution $p_\mathrm{N}$ and the target distribution $p_\mathrm{B}$ as a function of integration time $T$ for 10 trials (thin lines).
        The red dotted line shows convergence for a network of intrinsically stochastic (theoretically optimal) neurons that is guaranteed to converge to $p_\mathrm{B}$ for $T \rightarrow \infty$.
        (D) Kullback-Leibler divergence $\Dkl{p_\mathrm{N}}{p_\mathrm{B}}$ when sampling for $T=\unit[10^3]{s}$ from 100 different randomly generated target distributions.
      }
\end{figure}

The sampling process with networks of LIF neurons was examined in computer simulations.
Fig.~\ref{fig:3}A shows the spike pattern of a recurrent network of $K=5$ LIF neurons that sample from a randomly generated Boltzmann machine.
The parameters $b_k$ and $W_{kj}$ were drawn from the interval $[-0.6, 0.6]$. 
The thus defined target distribution $p_\mathrm{B}(\v z)$ is approximated by the distribution $p_N(\v z)$ of network states when the spike pattern is interpreted as samples $\v z^{(t)}$ by convolution with a $\tau_\mathrm{on} = \unit[10]{ms}$ box kernel.
Fig.~\ref{fig:3}B shows the average network distribution $p_\mathrm{N}(\v z)$ (blue bars) after $T=\unit[10]{s}$ simulation time alongside the target values $p_\mathrm{B}(\v z)$ (red lines) calculated from (\ref{eq:BM_joint}).
Sampled probabilities $p_\mathrm{N}(\v z)$ depict the mean over 10 independent simulation runs, errorbars reflect stochastic variations between individual runs.
The chosen integration time $T=\unit[10]{s}$ displays a conservative estimate of the maximum duration a neuronal ensemble experiences stable stimulus conditions in a behaving organism and can thus be expected to sample from a stationary distribution.
We find that the recurrent network of LIF neurons accurately encodes the target distribution over several orders of magnitude, within the precision imposed by the sample-based representation.

Fig.~\ref{fig:3}C shows how the network distribution becomes increasingly more reliable as more samples are considered.
After few samples, the network has generated a coarse approximation of $p_\mathrm{B}(\v z)$ that could serve as an ``educated guess'' in online computation tasks.
For simulation times $T$ well beyond biologically relevant timescales, systematic errors in $p_N(\v z)$ become apparent:
The KL divergence saturates on a non-zero value, while the (theoretically ideal) abstract model \cite{buesing2011neural} further converges towards the target distribution (red dotted line).
Fig.~\ref{fig:3}D shows that the sampling quality holds for a variety of similarly generated target distributions.

\begin{figure*}[tbp]
  \centering
  \includegraphics[]{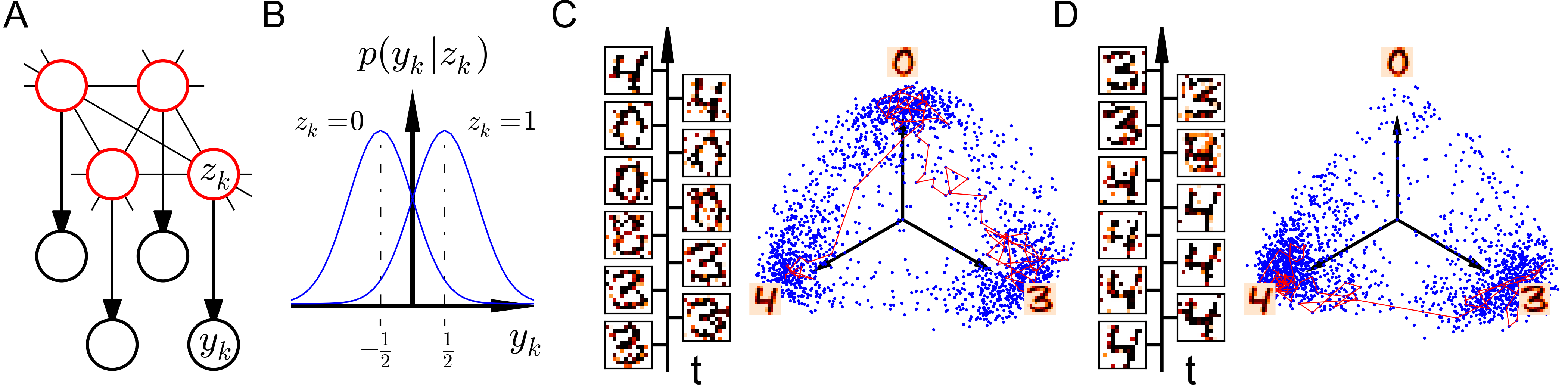}
  \caption{\label{fig:4}
      (A) Graphical model used for the probabilistic inference task.
      The network implements a Markov random field over latent variables $\v z$.
      Observables $y_k$ are conditionally independent given the network state.
      (B) A mixture-of-Gaussians likelihood model provides input to the sampling neurons.
      (C) Two dimensional projection of sampled states $\v z^{(t)}$ when sampling from the prior $p(\v z)$.
      The network preferentially spends time in modes close to the stored hand-written digits $0$, $3$, $4$.
      Solid line: network trajectory over $\unit[200]{ms}$.
      Color maps: Marginals $\bar z_k(t)$ averaged over $\unit[20]{ms}$.
      The time arrow covers the duration of the red trajectory and consecutive snapshots are $\unit[20]{ms}$ apart.
      (D) As in (C) when sampling from the posterior $\cp{\v z}{\v y}$ with incomplete observations $\v y$.
      The provided input $\v y$ is incompatible with digit $0$ and ambiguous with respect to digits $3$ and $4$.
      }
\end{figure*}

\subsection{Demonstration of probabilistic inference}

We conclude our investigation of sampling in recurrent networks of LIF neurons with an example of Bayesian inference based on incomplete observations.
A fully connected Boltzmann machine of $K=144$ neurons, aligned on a $12$x$12$ grid, was trained as an associative network \cite{Hinton26051995} to store three prototypic patterns, namely hand-written digits $0$, $3$ and $4$.
In the network, each cell is assigned to one pixel of the image.
Statistical correlations between the pixels as well as their mean intensities were encoded in the weights $W_{kj}$ and biases $b_k$ of the corresponding joint distribution $p(\v z)$.
This distribution reflects ``prior knowledge'' imprinted in the network.

The probabilistic model is augmented by adding real-valued input channels for each pixel.
Like the network variables $z_k$, input channels are associated with random variables $y_k \in {\mathbb R}, \, 1 \le k \le K$.
The resulting graphical model is sketched in Fig.~\ref{fig:4}A and entails the following structure for a full probabilistic model:
\begin{align}
  p(\v y,\, \v z) &= p(\v z) \cdot \prod_{k=1}^{K} \cp{y_k}{z_k} \quad .
\end{align}
The full model $p(\v y,\,\v z)$ connects the network variables $z_k$ of the prior to inputs $y_k$ by means of the likelihood $\cp{y_k}{z_k}$.
We have chosen a Gaussian likelihood with unit variance (see Fig.~\ref{fig:4}B):
\begin{align}
  \cp{y_k}{z_k} = \normal[\mu=z_k-\frac12,\,\sigma^2=1]{y_k} \;\;.
\end{align}
The task for the network is to implement the posterior distribution that follows from Bayes' rule: $\cp{\v z}{\v y} \propto p(\v z) \cdot \cp{\v y}{\v z}$.
The posterior combines two sources of information:
The likelihood $\cp{y_k}{z_k}$ tends to align the network state with the observation, i.e.\ $z_k=1$ for $y_k > 0$,
while the prior $p(\v z)$ reconciles the observations with knowledge on consistent activation patterns $\v z$.
In this way, the posterior $\cp{\v z}{\v y}$ evaluates all possible outcomes $\v z \,|\, \v y$ simultaneously by assigning a belief to each of them, and thus captures the model's (un-)certainty about different solutions.
A short derivation shows that the posterior $\cp{\v z}{\v y}$ is a Boltzmann machine for any input $\v y$, thus being compatible with the sampling dynamics of spiking networks. More specifically, we obtain the following abstract membrane potential:
\begin{align}
  v_k &= b_k + y_k + \sum_{j} W_{kj} \, z_j \;\;.
\end{align}
In the LIF domain, the sum $b_k + y_k$ is equivalent to an effective bias (\ref{eqn:biastrans}) and corresponds to an external current $\Iext_k = I^b_k + I^y_k$ that shifts $\bar u_k$ appropriately.
Thus, a network neuron receives synaptic input from recurrent connections and noise sources, as well as an external current, i.e.,  $I_k = \Irec_k + \Inoise_k + I^b_k + I^y_k$.

In case of $I^y_k = 0\;\forall k$, the network samples from the prior distribution $p(\v z) = \cp{\v z}{\v y = \v 0}$.
A two-dimensional projection of network states $\v z^{(t)} \sim p(\v z)$ is shown in Fig.~\ref{fig:4}C.
The sampled distribution exhibits three distinct peaks that correspond to the three hand-written digits stored in the recurrent weight matrix.
A closer look at the network trajectory reveals how the system stays in one mode for some duration, traverses the state space and then samples from a different mode of the distribution.
These dynamics also reflect in the marginals of the network variables under a $\unit[20]{ms}$ box filter,
$\bar z_k(t) = \frac{1}{\unit[20]{ms}}\int_{t-\unit[20]{ms}}^t z_k^{(t')} \, dt'$, shown in the color maps.

A typical scenario of stochastic inference on incomplete observations is shown in Fig.~\ref{fig:4}D.
Four input channels, located at the center of the grid, were picked to inject positive currents $I^y_k > 0$ to the network while all other inputs remained uninformative.
Positive currents $I^y_k$ encode positive values of the respective input pixels $y_k$ and were chosen such that the observation appeared incompatible with the digit 0, and remained ambiguous with respect to digits 3 and 4.
Accordingly, in the posterior distribution $\cp{\v z}{\v y}$ the $0$-mode is significantly suppressed while uncertainty about the provided cue is expressed by two distinct modes in the $3$ and $4$ directions.

\subsection{Discussion}

We have shown how recurrent networks of deterministic neurons in a spiking noisy environment can perform probabilistic inference through sampling from a well-defined posterior distribution.
Our approach builds on theoretical work by Buesing et al.~\cite{buesing2011neural} and extends Bayesian spiking network implementations to deterministic neuron models widely used in computational neuroscience.
For the analytical derivation and the computer simulations we have employed leaky integrate-and-fire neurons with conductance-based synapses.
However, the analysis can be readily transferred to other neuron models \cite{brette2005adaptive}.
The essential diffusion approximation relies on high-frequency spiking inputs that could be provided by the surrounding network and lead to strong synaptic conductances and fast membrane dynamics. 
Thereby, our derivation identifies a potential functional role of biologically observed high-conductance states within a normative framework of brain computation.

For mathematical tractability, simplifying modeling assumptions had to be made.
The neuron model only uses an absolute refractory time $\tauref$, which matches the activation time constant $\tauon$, and neglects any additional gradual recovery effects after a spike.
On the network level, we have assumed statistically independent noise sources and instantaneous axonal transmission.
Furthermore, post-synaptic potentials mediated through conductance-based synapses differ from the theoretically optimal rectangular shape and can lead to deviations from the target distribution outside of the high noise regime \cite{gerstner1992associative}.
However, computer simulations indicate that in most scenarios the above approximations are not critical.

Beyond neuroscience, the ability to perform probabilistic inference with deterministic neurons displays a promising computing paradigm for neuromorphic hardware systems.
Originally designed as neuroscientific modeling tools, these systems typically implement a physical model of integrate-and-fire neurons \cite{mitra2009real,Bruederle2011263}, which renders the application of the proposed networks straightforward.
In particular, the distributed nature of the sampling algorithm allows to exploit the inherent parallelism of neuromorphic architectures, fostering an application of neuromorphic hardware to online data evaluation and robotics.

\bibliography{refs}

\end{document}